\documentclass[3p]{elsarticle}

\usepackage{amssymb}
\usepackage{stackengine}
\usepackage[utf8]{inputenc}
 \usepackage{mathtools}
\usepackage{amsmath}
\usepackage{tikz}

 \usepackage{bbding}
\usepackage{pifont}
\def\abs{\@ifstar{\oldabs}{\oldabs*}}
\usepackage{rotating}

\usepackage{nomencl}
\usepackage{tgcursor}
\usepackage{amsmath,amsfonts,mathtools}
\usepackage{cite}
\usepackage{ wasysym }
\usepackage{multirow}
\usepackage{breqn}
\usepackage{array}
\usepackage{colortbl}
\usepackage{algorithmic}
\usepackage{mdframed}
\usepackage{longtable}
\usepackage{stackengine}
\usepackage{blindtext}
\usepackage{tikz}
\usetikzlibrary{shapes.geometric,backgrounds,fit,arrows}
\usepackage[utf8]{inputenc}
 \usepackage{mathtools}
\usepackage{amssymb}
\usepackage{amsmath}
\usepackage{amsthm}

\usepackage[ruled,vlined]{algorithm2e}
\def\abs{\@ifstar{\oldabs}{\oldabs*}}
\usepackage[ruled,vlined]{algorithm2e}
\def\abs{\@ifstar{\oldabs}{\oldabs*}}
\usepackage{nomencl}
\usepackage{caption}
\usepackage{subcaption}

\usepackage {hyperref}
\hypersetup{ colorlinks=true, % false: boxed links; true: colored links
linkcolor=blue, % color of internal links
citecolor=blue, % color of links to bibliography
filecolor=magenta, % color of file links
urlcolor=blue, bookmarksdepth=4 }

\definecolor{shadecolor}{rgb}{1,0.8,0.3}
\definecolor{lightgray}{rgb}{0.9,0.9,0.9}
\definecolor{yellow}{rgb}{1,0.5,0}

\usepackage{hyperref}
\hypersetup{ colorlinks = false, linkbordercolor = {1 0 0}
}

\usepackage{framed} % Framing content

\usepackage{multicol} % Multiple columns environment

\usepackage{nomencl} % Nomenclature package

\makenomenclature

\renewcommand*\nompreamble{\begin{multicols}{2}}

\renewcommand*\nompostamble{\end{multicols}}

\usepackage{etoolbox}
\renewcommand\nomgroup[1]{%
  \item[\bfseries \ifstrequal{#1}{A}{\textit{Abbreviations}}{
  \ifstrequal{#1}{B}{\textit{Indices and sets}}{
  \ifstrequal{#1}{C}{\textit{Parameters in the first stage}}{
  \ifstrequal{#1}{D}{\textit{Variables in the first stage}}{ }}}} ]}
\usepackage{accents}

\usepackage{blkarray}

\newcommand{\coo}{\ensuremath{\mathrm{CO_2}}}

\journal{International Journal of Electrical Power \& Energy Systems}

\begin{document}
% \begin{highlights} \item A framework for energy scheduling and trading of
% multiple energy hubs is proposed. \item Local markets for trading different
% forms of energy between energy hubs are designed. \item The stochastic energy
% scheduling problem for a holistic model of energy hub is presented. \item
% Energy hubs offering in local markets is formulated. \end{highlights}

\begin{frontmatter}

\title{Long-Term Benefits of Network Boosters for\\Renewables Integration and Corrective Grid Security}

\author[1]{Amin Shokri Gazafroudi\corref{cor1}}
\ead{shokri.g.amin@gmail.com}%\fnref{fn1}}
\cortext[cor1]{Corresponding author}

\author[2]{Elisabeth Zeyen}
\ead{e.zeyen@tu-berlin.de }

\author[1]{Martha Frysztacki}
\ead{martha.frysztacki@kit.edu}

\author[2]{Fabian Neumann}
\ead{f.neumann@tu-berlin.de}

\author[2]{Tom Brown}
\ead{t.brown@tu-berlin.de}

\address[1]{Institute for Automation and Applied Informatics, Karlsruhe Institute of Technology (KIT), 76131 Karlsruhe, Germany}

\address[2]{Institute of Energy Technology, Technical University of Berlin, Einsteinufer 25, 10587 Berlin, Germany}

\begin{abstract}

The preventative strategies for $\mathcal{N}-1$ network security dominant in
  European networks mean that network capacity is kept free in case a line
  fails.  If instead fast corrective actions are used to overcome network
  overloading when single lines fail, this  has the potential to free up network
  capacity that is otherwise underused in preventive $\mathcal{N}-1$ security
  strategies. In this paper, we investigate the impact on renewable integration
  of a corrective network security strategy, whereby storage or other
  flexibility assets are used to correct overloading shortly after line outages.
  In this way, we find significant cost savings for the integration of renewable
  energy of up to 2.4 billion euros per year in a aggregated 50-bus model of the
  German power system utilizing these flexibility assets, so-called
  \textit{network boosters} (NB). This offers a role for storage beyond energy
  arbitrage or ancillary services like frequency control. While previous
  literature has focused on the  potential savings of NB in the short-term
  operation, we focus on the long-term  benefits in systems with high shares of
  renewable energy sources, where the  capacities and dispatch of generation and
  NB are optimised. We demonstrate the  benefits of NB for various shares of
  renewable energy, NB and flexibility  costs, as well as different allowed
  levels of temporary overloading the lines  in both (i) a sequential model,
  where long-run generation investments are  optimised separately from the NB
  capacities, and (ii) a simultaneous model,  where generation is co-optimised
  with NB investment so that mixed preventive-corrective approaches are
  possible.

\end{abstract}

%%Graphical abstract \begin{graphicalabstract} %\includegraphics{grabs}


\begin{keyword}
Contingency \sep flexibility \sep network booster \sep power transmission system
\sep reliability \sep security-constrained optimal power flow.



%% MSC codes here, in the form: \MSC code \sep code % or \MSC[2008] code \sep
%code (2000 is the default)

\end{keyword}

\end{frontmatter}

% \begin{scriptsize} \printnomenclature[1.5cm] \end{scriptsize}

\begin{footnotesize}
\printnomenclature[1.5cm]
\end{footnotesize}

\setlength{\nomlabelwidth}{1.5cm}	
% \nomenclature[A,01]{DSO}{ Distribution System Operator}

% %index  
% \nomenclature[B,01]{$j,j^{'}$}{ Buses} 

% %FIRST STAGE %parameters \nomenclature[C,01]{$\lambda_{jt\omega}^{lm}$}{
% Predicted energy trading price in the LEM for prosumer $j$ at time slot $t$
% and scenario $\omega$}

% %variables \nomenclature[D,01]{$OF^{p}_{j}$}{ Objective function for prosumer
% $j$ }

\section{Introduction}

\subsection{Background}

Traditional preventive approaches to the $\mathcal{N}-1$-secure operation of
transmission networks keep power flows in transmission lines below line
capacities so that if a line fails, no other line in the network becomes
overloaded. This leads to the under-utilization of network assets and hinders
the economic integration of generation sources, particularly those like wind and
solar that are transported over long distances. In Europe, such preventative
approaches dominate strategies for $\mathcal{N}-1$ network security. An
alternative corrective approach is to react to the line failure only after it
happens, by activating flexibility resources such as storage or demand-side
management to counter-act any overloading that results in the rest of the
network \citep {zhao2015unified}. Thus, flexibility assets act within minutes to
prevent long-term overloading. This corrective approach allows a higher
utilization of the transmission network, and could benefit the integration of
renewable energy sources (RESs), particularly in regions where transmission
expansion is slow because of public acceptance problems. In addition, corrective
network security offers an additional revenue stream to assets like batteries,
besides regular energy arbitrage and ancillary services such as reserves.

\subsection{Literature review}

Several studies have investigated the role of contingency constraints as well as
flexibility resources in power transmission systems. For instance, operational
flexibility and $\mathcal{N}-1$ constraints for transmission lines were
considered in a combined generation and transmission planning approach in \citep
{wang2021transmission}. Authors in \citep {bhardwaj2021risk} applied a
non-linear security constrained optimal power flow (SCOPF) model with mixed
preventive-corrective contingency measures, finding a role for both preventive
and corrective actions. Authors in \citep {bucher2015managing} discussed a
framework to employ operational flexibility among transmission system operators
(TSOs) in multi-area power systems subject to $\mathcal{N}-1$ security criteria.
Authors in \citep {ramesh2020reducing} proposed a corrective approach to solve a
$\mathcal{N}-1$ security-constrained unit commitment problem to minimize
congestion of transmission lines and curtailment of RESs. Authors in \citep
{zhang2020corrective} presented a corrective approach to avoid line overloading.
In \citep {nikoobakht2019flexible}, re-dispatching of flexible resources is used
in the context of SCOPF as a corrective strategy in response to line outages. A
preventive approach has been utilized by authors in \citep
{gazafroudi2021topology}, where $\mathcal{N}-1$ security conditions have been
approximated based on the network's topology by analysing the polytope of
feasible nodal net power injections under contingency constraints to reduce the
problem's computational burden. Similarly, authors in \citep {weinhold2020fast}
introduced a geometric algorithm based on the feasible space of net power
injections for reducing the number of preventive contingency constraints.

While the mathematical formulation of preventive and corrective
security-constrained optimal power flow is well understood \citep
{capitanescu2011state,van2018modeling}, most investigations of the benefits of
corrective security have focused on short-term cost reductions in the operation
of the power system where the investments in generation and flexibility assets
are fixed. For example, it was shown in \citep {strbac1998method} that
corrective actions can significantly reduce the costs of maintaining operational
security in a case study with the 24-bus IEEE reliability test system. Benefits
of a corrective approach were also found in a European case study that showed
that investment in corrective controls was complementary to transmission
expansion \citep {muller2013techno}, but generation capacities were once again
exogenous. High share of renewable generation were considered in a multi-period
stochastic context with both preventative and corrective network security in
\citep {ALIZADEH2022107992}, but the impact on generation planning was not
considered. In the German context, several recent studies have shown that
corrective measures could lead to a significant reduction in congestion
management measures such as redispatch \citep
{hoffrichter2019simulation,kollenda2020}. Transmission-connected battery systems
that perform these corrective actions have been called \textit{network boosters}
(NB) by the German transmission system operators \citep {nep}. In \citep
{KOLSTER2020115870} it was shown that there are significant distributed
flexibility potentials on the demand side too, including up to 16~GW of upward
flexibility potentials from power-to-heat units in district heating networks.

\subsection{Contributions}
While previous research has included storage capacities in generation planning
for energy arbitrage between time periods or ancillary services \citep
{qiu2016stochastic, joubert2018impact, frysztacki2021strong, fiorini2017sizing},
we propose a planning model that for the first time uses the storage and
flexible systems as NB resources for corrective network security. This approach
avoids having to maintain free network capacity preventatively and thereby
allows lines to be fully loaded. Additionally, we explore the longer-term impact
of a corrective approach to $\mathcal{N}-1$ security on the distribution of
generation capacities within Germany in this paper. By freeing up grid capacity,
network boosters allow wind and solar generation to be placed at sites with
better capacity factors, thus reducing the overall costs of the power system.
This impact is expected to be particularly high in countries like Germany, where
network expansion is known to experience long delays \citep {tyndp}, putting
pressure on grid operators to maximise the use of existing assets. We explore
the benefit as a function of different levels of renewable energy, as a function
of the network booster cost and as a function of the allowed temporary
overloading of the lines before the network booster can be activated.

Our main contributions are:

\begin{itemize}
  \item to present a model that co-optimizes network booster placement with
  generation investment;
  \item to explore the difference between an approach where generation and
  network booster capacities are optimized sequentially versus a simultaneous
  co-optimization that allows for mixed preventive-corrective security;
  \item to explore the dependency of network booster benefit on \coo\ reduction
  targets, which are a proxy for renewable penetration in the German context;
  \item to analyse the effect of the cost of network boosters (particularly
  relevant given the uncertainty of battery costs and the possible substitution
  of batteries with demand-side management);
  \item to investigate the sensitivity to the temporarily admissible
  transmission loading (TATL) of transmission lines.
\end{itemize}

Section \ref{sec:nb} introduces the modelling of NB. In Section
\ref{sec:strategies}, TATL strategies for the NB are presented. Then, the
performance of our proposed models are studied in Section \ref{sec:simulation}.
We draw attention to the limitations of the modelling in Section
\ref{sec:limitations}. Finally, our findings are concluded in Section
\ref{sec:conclusion}.

\section{Network Booster Modelling} \label{sec:nb}

\begin{figure}[!t]
  \centering
  \includegraphics[width=0.6\linewidth]{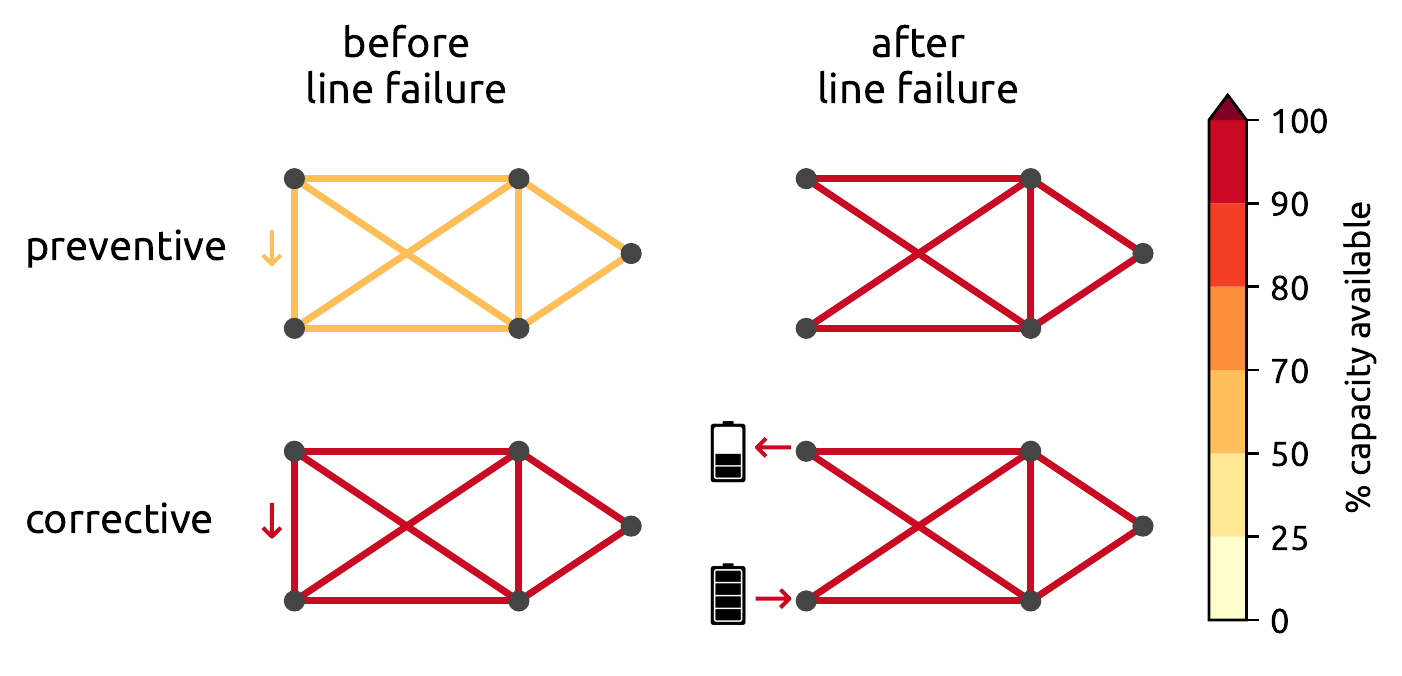}
  \caption{Illustration of the concept of preventive strategies to maintain
  network security under $\mathcal{N}-1$ conditions compared to network boosters
  as a corrective $\mathcal{N}-1$ security strategy.}
  \label{fig:nb_flow}
\end{figure}

First, we introduce our generalized network booster (NB) model to investigate
possible benefits of a corrective $\mathcal{N}-1$ strategy compared to
preventative $\mathcal{N}-1$ security. The network boosters work by injecting or
withdrawing power from the grid within minutes after a line fails, so that no
other line becomes overloaded. Lines may be temporarily overloaded for the few
minutes it takes the network booster to come into operation.

The objective of the mathematical problem is to minimize the total required NB
capacity at all buses so that we can retain the security of the network
considering all single line failures as illustrated in Fig. \ref{fig:nb_flow}.
The proposed NB problem optimizes the capacity to adjust power or demand.
$P^+_i$  represents the upward NB capacity to increase power or reduce demand at
bus $i$,  $P^-_i$ the downward NB capacity to decrease power or increase demand
at bus $i$. Thus, both upward and downward NB capacities are positive:
\begin{equation}
  \label{eq:nb5}
  P^+_i  \geq 0  , \quad\forall  i,
\end{equation}
\begin{equation}
  \label{eq:nb6}
  P^-_i  \geq 0 , \quad \forall i.
\end{equation}

The dispatch variables of the NB resources $p^+_{i,t,\ell}$ and $p^-_{i,t,\ell}$
are defined at bus $i$ in time $t$ for the outage of line $\ell$ to be positive
and  limited to the upward and downward NB capacities respectively:
\begin{equation}
  \label{eq:nb7}
  0 \leq p^+_{i,t,\ell}  \leq P^+_i   , \quad\forall i,\forall t,\forall\ell.
\end{equation}
\begin{equation}
  \label{eq:nb8}
  0 \leq p^-_{i,t,\ell}  \leq P^-_i  , \quad\forall i,\forall t,\forall\ell.
\end{equation}

The total power injected into the network by all NB resources is equal to the
total power absorbed from the network from NB resources, assuming a lossless
transmission network:
\begin{equation}
  \label{eq:nb9}
  \sum_i \left(p^+_{i,t,\ell} - p^-_{i,t,\ell}\right)  = 0  , \quad \forall t, \forall\ell.
\end{equation}

Finally, the minimum and maximum constraints for avoiding overloading in
transmission line $\ell$ when line $k$ fails by operating NB resources under
$\mathcal{N}-1$ security condition is given by:
\begin{flalign}
  \label{eq:nb10}
  -F_\ell & \leq f_{\ell,t} + LODF_{\ell k} f_{k,t} \\\nonumber & + \sum_i
          \left[(PTDF_{\ell i}+ LODF_{\ell k} PTDF_{k i}) (p^+_{i,t,k} -
          p^-_{i,t,k})\right]             \\\nonumber & \leq F_\ell ,
          \qquad\forall t, \forall k, \forall \ell \neq k,
\end{flalign}

where $f_{\ell,t}$ represents the power flow in line $\ell$ at time $t$ before
the outage of line $k$ and $F_\ell$ is the nominal power capacity of
transmission line $\ell$, also called the \textit{Permanently Admissible
Transmission Loading (PATL)}. The $PTDF$ matrix represents the \textit{Power
Transfer Distribution Factors} which relate nodal power injections to the line
flows \citep {wood2013power}. The $LODF$ matrix introduces the \textit{Line
Outage Distribution Factors} \citep {Guo09,hinojosa2017stochastic} that measure
the sensitivity of line flows to the outage of other lines
\begin{equation}
  \label{eq:lodf}
  LODF_{\ell,k}= \frac{[PTDF \cdot K]_{\ell,k}}{1- [PTDF \cdot K]_{\ell,\ell}}. \quad\forall \ell,
\end{equation}
If $\ell=k$, we define $LODF_{\ell,k}=-1$. $K_{i,\ell}$ denotes the incidence
matrix of the transmission grid which has non-zero values +1 if line $\ell$
starts at bus $i$ and -1 if line $\ell$ ends at bus $i$. The orientation of the
lines is arbitrary but fixed.

The first two terms in Eq. \eqref{eq:nb10} give the power flow in line $\ell$
immediately after the outage of line $k$, while the sum computes the effect of
the network boosters on the power flow to counteract the effects of the outage.
It is assumed that it may take several minutes to active the network boosters.
In the meantime, we assume that the transmission lines can be loaded more than
their nominal power capacity, up to the so-called \textit{Temporarily Admissible
Transmission Loading (TATL)}, before the NB resources activate. It is notable
that the steady state analysis and short-term dynamic analysis is not performed
in our proposed model. This way, the relation between the TATL $\tilde F_\ell$
and nominal power capacity (PATL) for line $\ell$ is given by:

\begin{equation} \label{eq:tatl_nomimal}
  \tilde F_\ell = f^{tatl} F_\ell , \quad\forall \ell,
\end{equation}

where $f^{tatl}$ is the TATL factor, which is larger than one ($f^{tatl}>1$). In
this way, if line $\ell$ fails, some lines can become temporarily overloaded,
until the NB resources act within a certain time, e.g. 2-5 minutes, to eliminate
the overloading. Typical TATL values are 10-30\% higher than the PATL \citep
{kollenda2020}. Since the time until the NBs act depends on the NB installation
and the time has an influence on the TATL, we provide a sensitivity analysis for
the TATL. In addition, the NB resources need to be able to operate for a certain
time, e.g. 30 minutes, to give time for re-dispatching measures such as reducing
wind feed-in or firing up gas power plants. To ensure that no line $\ell$ is
loaded above the TATL after  the failure of line $k$, additional constraints
must be enforced:
\begin{equation}\label{eq:tatl n-1}
  \left| f_{\ell,t} + LODF_{\ell k} f_{k,t} \right| \leq \tilde F_\ell \hspace{1cm} \forall  k,\ell \neq k,t
\end{equation}

\section{Network Booster Planning Approaches} \label{sec:strategies}

\begin{figure}[!t]
  \centering
  \includegraphics[width=0.6\linewidth]{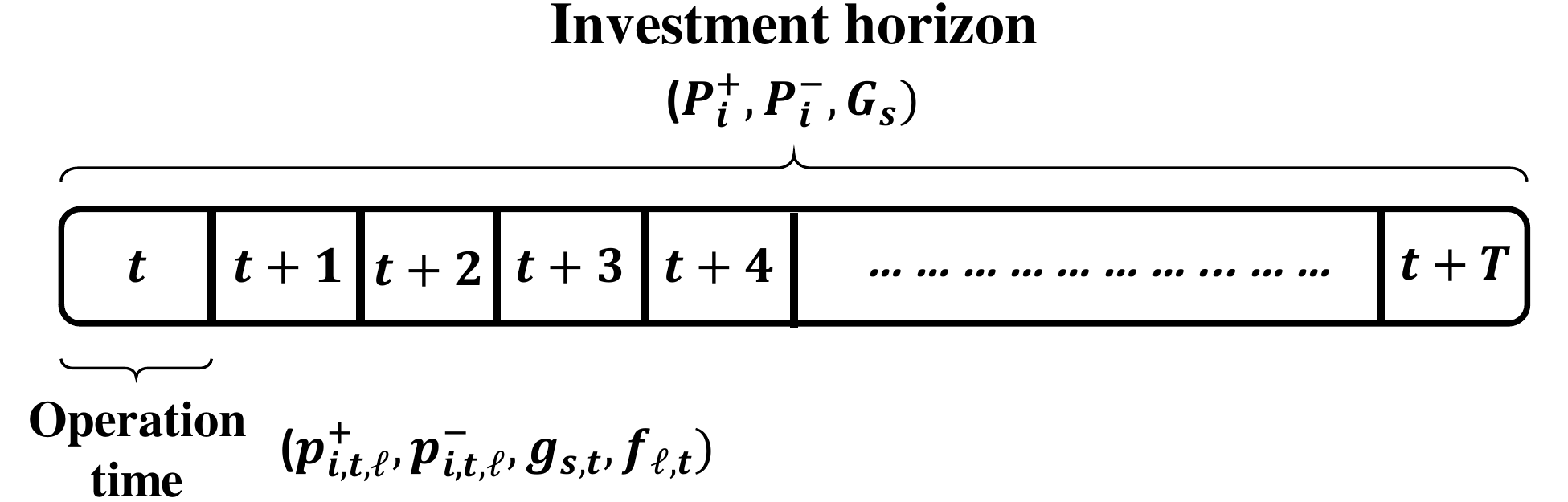}
  \caption{Operation and investment time horizons for the discussed network booster strategies.}
  \label{fig:horizon}
\end{figure}

In this section, we propose two strategies for optimizing the size and placement
of network boosters along with the long-term electricity generation fleet. The
first \textit{sequential} strategy separates the problem into a long-term
generation capacity expansion problem where it is assumed that network boosters
take care of $\mathcal{N}-1$ security correctively, followed by a separate
problem to determine the required investment in the network boosters. From a
market perspective, this could represent the separation of long-term generation
investment from the secure operation of the grid by the network operator. In the
simultaneous problem, investment in generation and network boosters is optimized
jointly, allowing mixed preventive-corrective strategies for $\mathcal{N}-1$
security. The related decision-making variables and their corresponding time
horizons are illustrated in Fig. \ref{fig:horizon}. Thus, $p^{+}_{i,t,\ell}$,
$p^{-}_{i,t,\ell}$, $g_{s,t}$ and $f_{\ell,t}$ are operation decision-making
variables for time $t$, and $P^{+}_{i}$, $P^{-}_{i}$ and $G_{s}$ are investment
decision-making variable over the studied time horizon. The combined network
booster and generation planning problem is computationally demanding (requiring
100~GB memory for the German case study), which motivates the linearisation of
all network and generation constraints.

\subsection{Sequential Model}

Our proposed sequential model splits into two stages. The first stage solves the
electricity generation investment problem, while the NB placement problem is
solved in the second stage. In this way, the power flow ($f_{\ell,t}$) obtained
from the investment problem is an input parameter of the NB placement problem.

\subsubsection{First Stage -- Electricity Generation Investment Problem}

The objective function for the electricity generation investment problem is to
minimise annual system costs, composed of the capital and operation costs of the
generation fleet:
\begin{equation}
  \label{eq:lopf_obj}
  \min_{G_{s},g_{s,t},f_{\ell,t}} \sum_s  c_s G_s +\sum_{s,t} o_s g_{s,t},
\end{equation}
\begin{equation}
  \hspace{1.35 cm} \text{subject to \eqref{eq:tatl_nomimal}, \eqref{eq:tatl n-1}, \eqref{eq:lopf_bal} - \eqref{eq:lopf_co2}},  \nonumber
\end{equation}
where the first and second terms represent the capital and operation costs for
generators, respectively. $G_s$ represents the nominal power of generator $s$
that can be built at an annual cost of $c_s$, and $g_{s,t}$ represents the
dispatched power of generator $s$ at time $t$, which is multiplied with the
marginal cost of operation $o_s$. As illustrated in Fig. \ref{fig:horizon},
$T+1$ periods for representing demand and weather conditions are used to
dimension the investment problem.

Furthermore, the investment problem is subject to  some techno-economic and
physical constraints which are presented in the following. Eq.
\eqref{eq:lopf_bal} presents the balancing constraints, which require
generation, demand and network flows to match at any bus $i$ and at any time $t$
\begin{equation}\label{eq:lopf_bal}
  \sum_s K^g_{i,s}g_{s,t} - d_{i,t} = \sum_\ell K_{i,\ell}f_{\ell,t}, \quad \forall t, \forall i,
\end{equation}
where $K^g_{i,s}$ represents the incidence matrix of generators for mapping
generator $s$ at bus $i$. Thus, $K^g_{i,s}=1$ if generator $s$ is located at bus
$i$, otherwise $K^g_{i,s}=0$. $K_{i,\ell}$ denotes the incidence matrix of the
transmission grid, as previously noted.

The capacities of generation are constrained by a maximum installable potential
$\overline{G}_s$:
\begin{equation}
  \label{eq:lopf_gen1}
  0 \leq G_{s} \leq \overline{G}_s, \quad \forall s.
\end{equation}
Moreover, maximum and minimum limitations of dispatched power of generators are
expressed by
\begin{equation}
  \label{eq:lopf_gen2}
  0 \leq g_{s,t} \leq \overline{g}_{s,t}G_{s}, \quad \forall s, \forall t,
\end{equation}
where $\overline{g}_{s,t}$ represents the time-dependent availability factor for
power dispatch given in per unit of the generator’s capacity. This is relevant
for modelling variable renewables like wind and solar. Similarly, the absolute
power flow in lines is constrained to their permanent transmission line
capacities (PATL):
\begin{equation}\label{eq:lopf_line1}
  - F_\ell \leq f_{\ell,t} \leq F_\ell, \quad  \forall \ell, \forall t.
\end{equation}

Furthermore, the model features a constraint that limits the CO$_2$ emissions to
a desired target level, $\Gamma_{\coo}$.
\begin{equation}\label{eq:lopf_co2}
  \sum_{s,t} \frac{e_s}{\eta_s} g_{s,t} \leq \Gamma_{\coo},
\end{equation}
where $e_s$ denotes the specific emissions of the fuel, $\eta_s$ is the
generator efficiency. This model formulation classifies as linear problem (LP).

\begin{figure}[!t]
  \centering
  \includegraphics[width=0.6\linewidth]{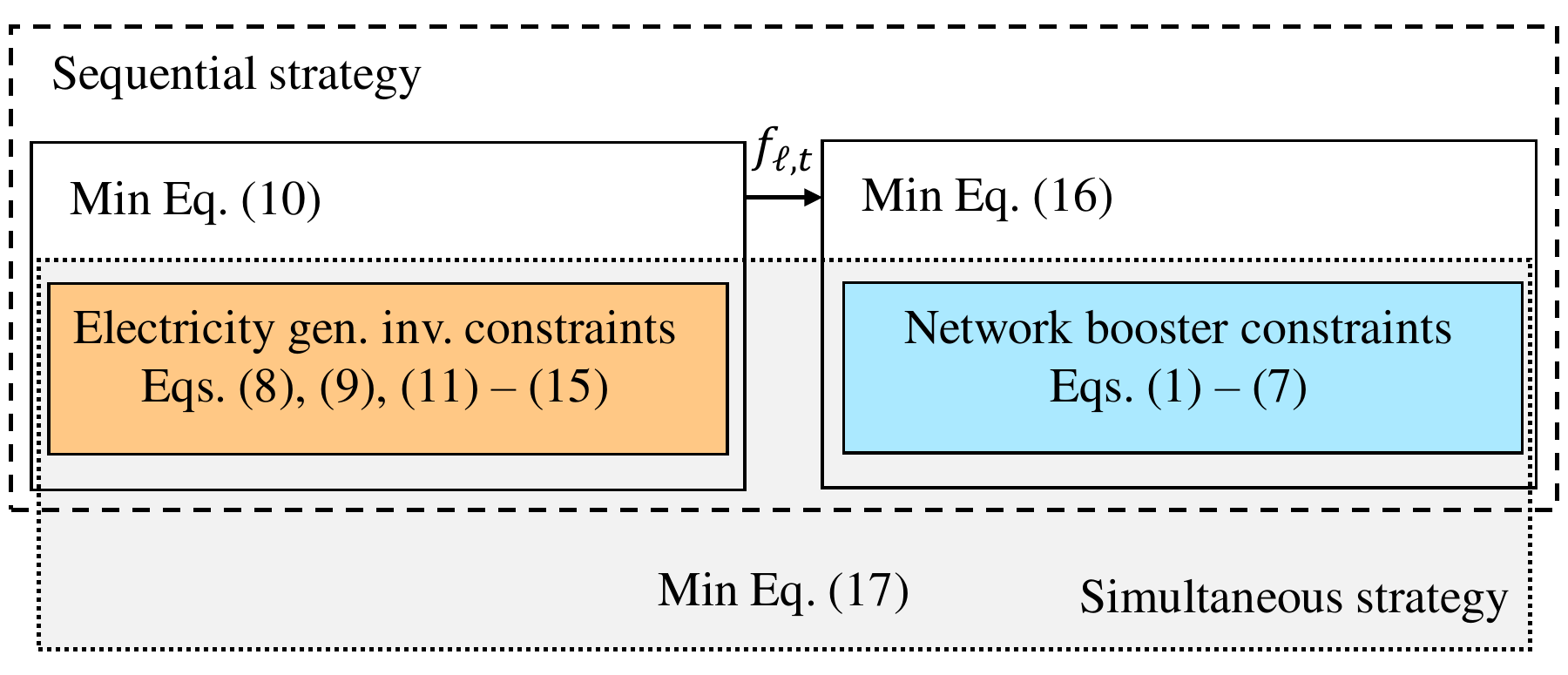}
  \caption{Problem formulations for the proposed sequential and simultaneous network booster strategies.}
  \label{fig:strategies}
\end{figure}

\subsubsection{Second Stage -- \textit{Network Booster} Problem}

In the second stage of the sequential model, the NB placement problem is solved
globally based on the planned generation fleet and operation considering all
single line outages. With the equations from the generalized NB model as
described in Section \ref{sec:nb}, we seek to minimize the total cost of network
booster investment and operation:
\begin{flalign}\label{eq:nb_obj}
  \min_{p^+_{i,t,\ell},p^-_{i,t,\ell},P^+_i,P^-_i} & \sum_i \left( c^+P^+_i +
  c^-P^-_i\right)                               \\\nonumber + & \sum_{i,t,\ell}
  \left( o^+ p^+_{i,t,\ell} + o^- p^-_{i,t,\ell}\right),
\end{flalign}
\begin{equation}
  \hspace{1.35 cm} \text{subject to \eqref{eq:nb5} - \eqref{eq:lodf}}, \nonumber
\end{equation}
where $c^+$ ($c^-$) represents cost for upward (downward) NB capacity. The
coefficients $o^+$ and $o^-$ represent the compensation cost paid to the NB
resources, i.e. demand side management  consumers, or running costs for energy
storage system, for dispatching when lines failed in the transmission network.
Moreover, it is is important to note that $f_{\ell,t}$ is a model input rather
than a decision variable in this second stage. Values for $f_{\ell,t}$ are taken
from the first stage and used as an exogenous parameter of the second stage in
the sequential NB problem. This model formulation is a linear problem (LP) as
well.

\subsection{Simultaneous Model}

Our proposed simultaneous model for the NB problem is an alternative to the
sequential model. In this simultaneous model, the electricity generation
investment problem and the generalized NB problem are co-optimized in a combined
linear problem (LP). Thus, the objective function for TATL simultaneous NB
problem is represented by:
\begin{flalign}\label{eq:nb_obj_sim} &
   \min_{G_{s},g_{s,t},f_{\ell,t},p^+_{i,t,\ell},p^-_{i,t,\ell},P^+_i,P^-_i}
   \sum_s  c_s G_s +\sum_{s,t} o_s g_{s,t}   \\\nonumber & + \sum_i \left(
   c^+P^+_i + c^-P^-_i\right) + \sum_{i,t,\ell} \left( o^+ p^+_{i,t,\ell} + o^-
   p^-_{i,t,\ell}\right),
\end{flalign}
\begin{equation}
  \hspace{0.1 cm} \text{subject to \eqref{eq:nb5} - \eqref{eq:tatl n-1}}, \eqref{eq:lopf_bal} - \eqref{eq:lopf_co2}, \nonumber
\end{equation}
where the power flow, $f_{\ell,t}$, is a decision-making variable in the
simultaneous NB problem.  The simultaneous model allows the decision-maker to
choose between preventive and corrective measures for each line. If the
decision-maker decides against a corrective strategy, it can set $p^+_{i,t,\ell}
= p^-_{i,t,\ell} = 0$ and the Eq. \eqref{eq:nb10} reverts to the standard
preventive $\mathcal{N}-1$ constraint for an outage of line $k$. This
endogeneous choice between corrective and preventive strategies cannot be
represented by the sequential model. The full problem formulations for
sequential and simultaneous network booster strategies are summarized in Fig.
\ref{fig:strategies}.

\section{Simulation Results}
\label{sec:simulation}

In this section, we study the performance of our proposed NB strategies in
finding cost-efficient ways to protect network operation against line outages
for a German test case, and observe where flexibility resources are optimally
placed. The optimisations are run through the open source modelling framework
PyPSA \citep {PyPSA}. The results are evaluated in a 50-bus German transmission
network using data from the open-source PyPSA-Eur model of the European
transmission power system, including time-series for load and renewable
availability for each region and existing power plants \citep {horsch2018pypsa}.
New investments are allowed at each location in wind, solar and fossil gas
generation capacities.

Moreover, one hundred typical hours are chosen to represent typical demand and
weather conditions. The hours are selected from historical data using the
$k$-means algorithm and weighted according to their frequency of occurrence
using the tsam library \citep {HOFFMANN-21-tsam}. Different numbers of typical
hours were initially investigated, but only above 50 hours were there enough
representative time periods for the objective function to be stable as the
temporal resolution was increased, indicating that the variability of wind and
solar was correctly captured. Thus, one hundred hours were chosen since this was
the upper limit for which we could solve the simultaneous model (the upper limit
being the memory usage of 100~GB). It is also assumed that a nodal pricing
regime is used for the dimensioning of generation assets.

The costs of the network boosters are based in the reference case on batteries
with an energy capacity cost of 142~\euro/kWh and a power capacity cost of
160~\euro/kW, assuming an energy capacity equal to 30 minutes at full power.
These cost assumptions are based on forecasts for the year 2030 from the Danish
Energy Agency \citep {dea2022}. The costs are annualized assuming a cost capital
of 7\% to give an annualized cost of $c^+ = c^- = C_f = 23$~\euro/kW/a (this
cost includes 0.5~MWh of energy storage for every 1~MW of power). These costs
are then varied in a sensitivity analysis. Very small operating costs of $o^+ =
o^-$ are assumed to avoid the batteries dispatching unless they are really
needed, and to avoid simultaneous charging and discharging. The batteries with
positive NB capacity are assumed to be fully charged before the outages, while
the batteries with negative NB capacity are assumed to be completely empty
before the outages.

The TATLs are uniformly assumed to be 30\% higher than the PATLs in the default
case, and are then varied in a sensitivity analysis. We study different levels
of \coo\ emission reduction levels compared to levels in the year 1990 in the
electricity sector.

\subsection{Sequential model}

\begin{figure}[!t]
  \begin{subfigure}{0.5\linewidth}
    \centering
    \caption{30\% \coo\ emission reduction}
    \includegraphics[width=\linewidth]{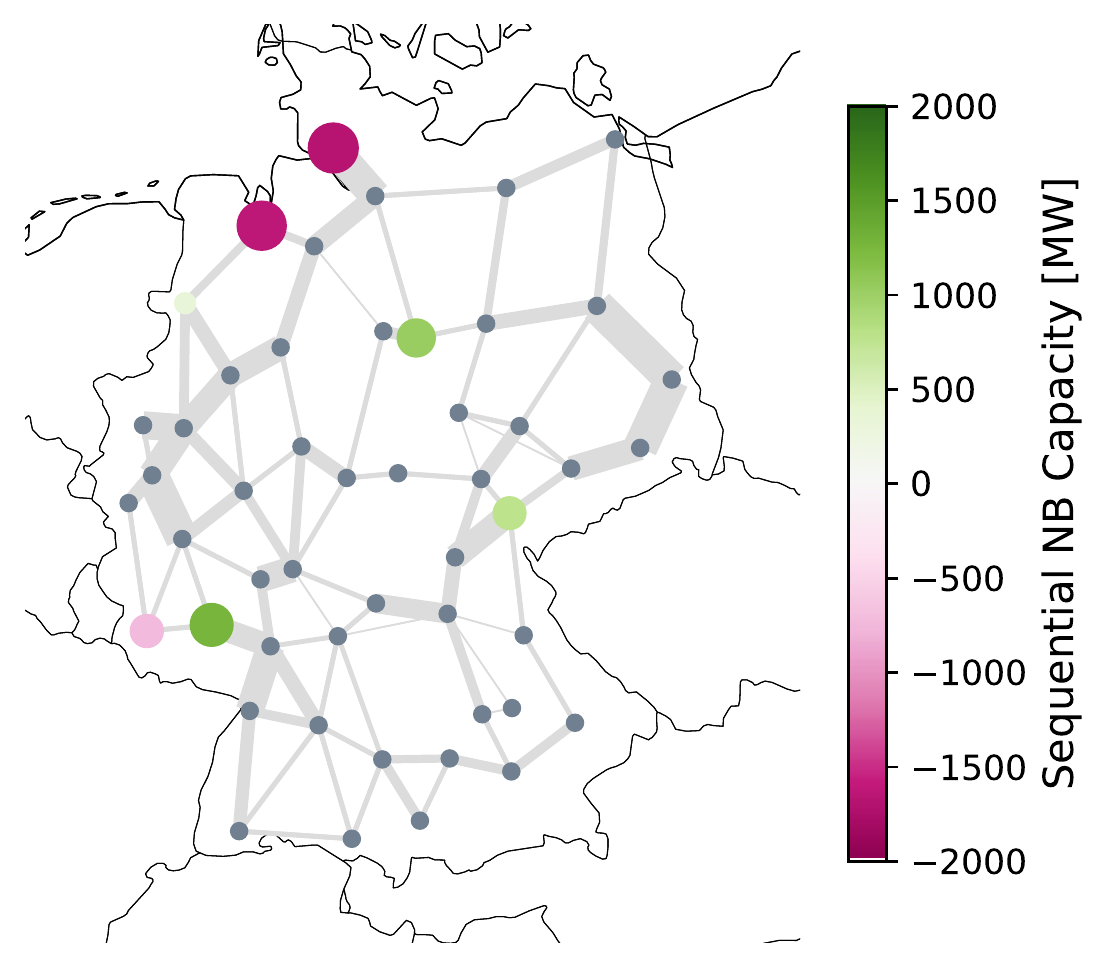}
  \end{subfigure}
  \begin{subfigure}{0.5\linewidth}
    \centering
    \caption{90\% \coo\ emission reduction}
    \includegraphics[width=\linewidth]{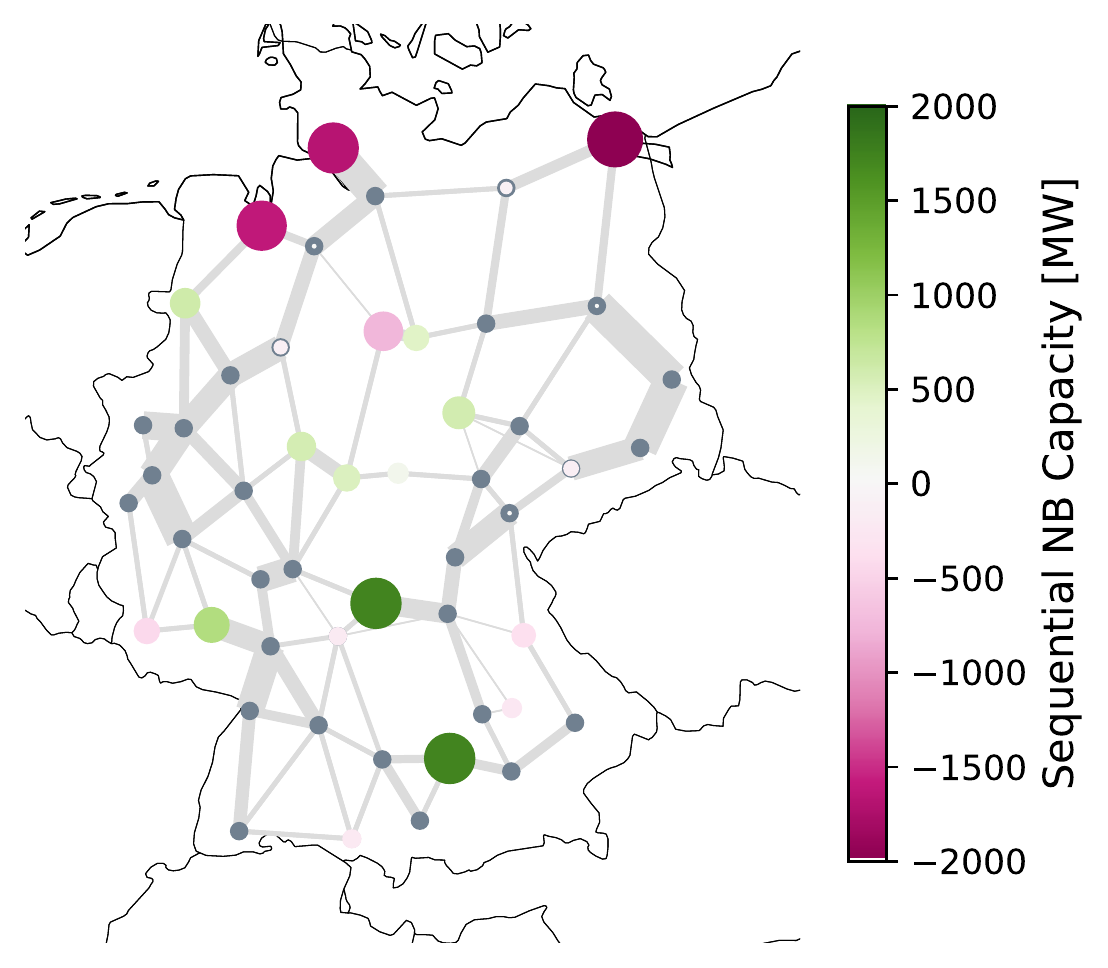}
  \end{subfigure}

  \caption{Impact of \coo\ emission reduction scenarios on map of allocated NB
  resources in the sequential model ($f^{tatl}$ = 1.3).}
  \label{fig:map_seq}
\end{figure}

\begin{figure}[!t]
  \begin{subfigure}{0.5\linewidth}
    \centering
    \caption{}
    \includegraphics[width=\linewidth]{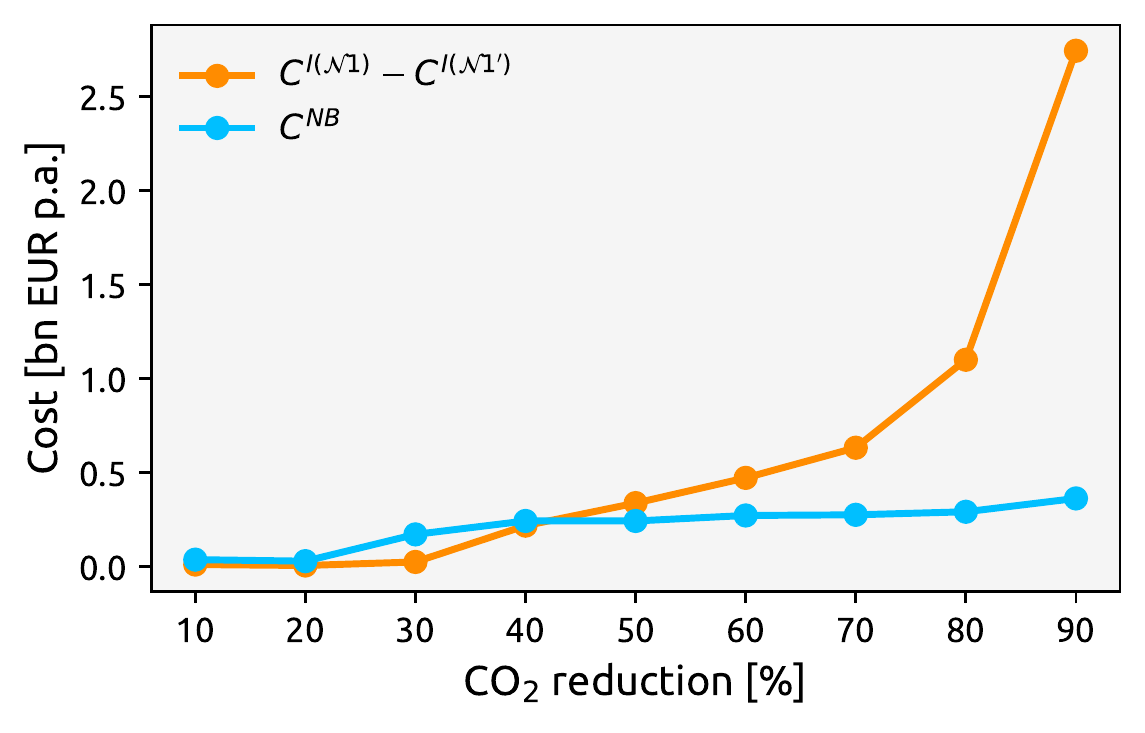}
  \end{subfigure}
  \begin{subfigure}{0.5\linewidth}
    \centering
    \caption{}
    \includegraphics[width=\linewidth]{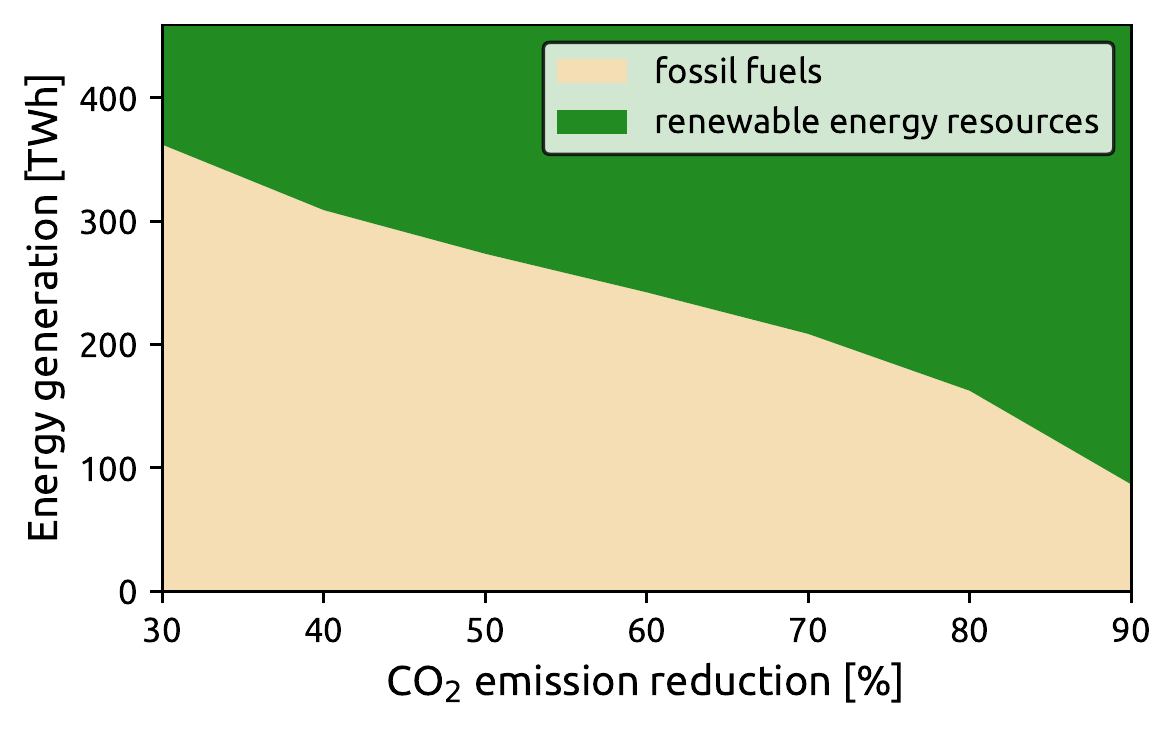}
  \end{subfigure}
  \caption{Impact of \coo\ emission reduction scenarios on (a) the NB cost
  ($C^{NB}$) and the difference between investment cost of full $\mathcal{N}-1$
  case and $\mathcal{N}-1'$ case with TATL ($C^{I({\mathcal{N}-1})}
  -C^{I(\mathcal{N}-1')}$) with $C_f=23$~\euro/kW/a, (b) energy generation mix
  in the sequential model.}
  \label{fig:carbon_seq}
\end{figure}

\begin{figure}[!t]
  \begin{subfigure}{0.5\linewidth}
    \centering
    \caption{30\% \coo\ emission reduction}
    \includegraphics[width=\linewidth]{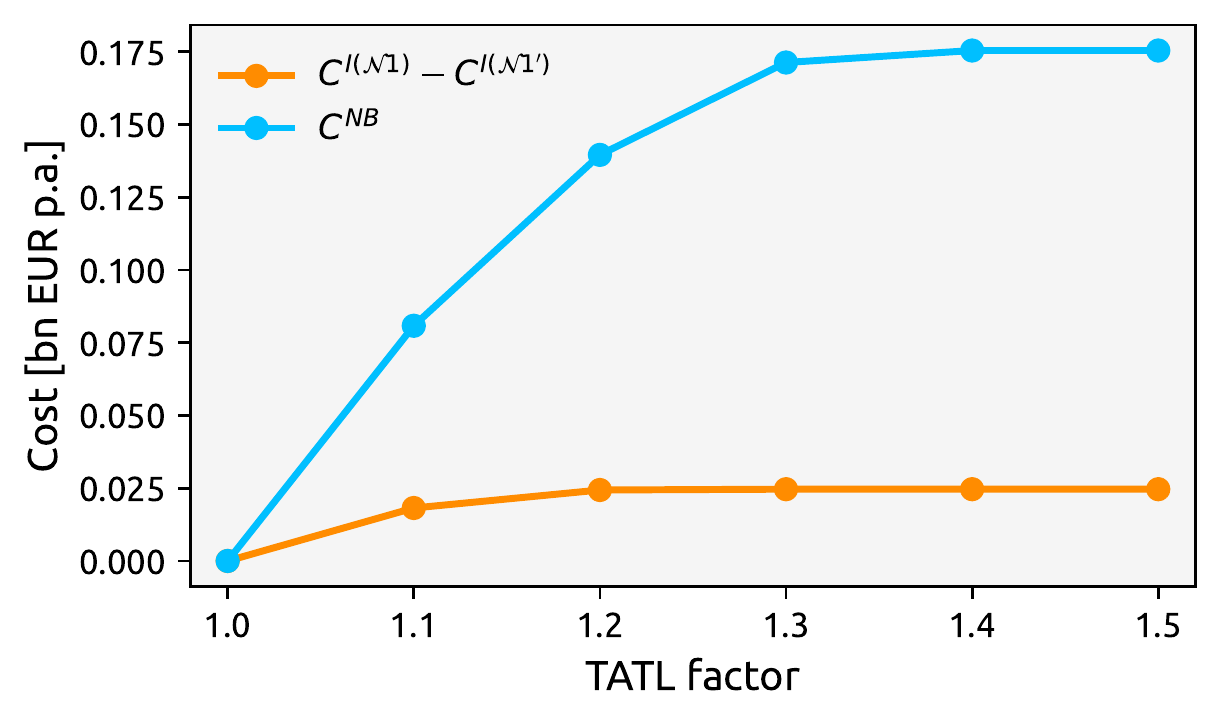}
  \end{subfigure}
  \begin{subfigure}{0.5\linewidth}
    \centering
    \caption{90\% \coo\ emission reduction}
    \includegraphics[width=\linewidth]{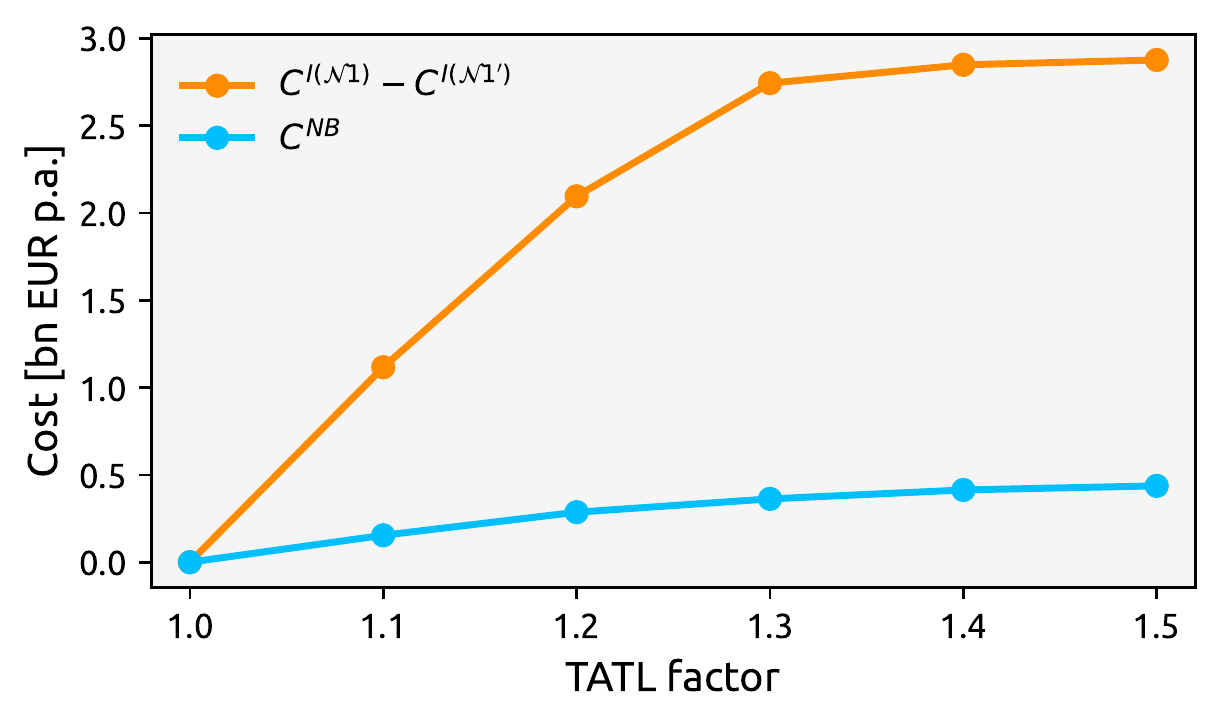}
  \end{subfigure}
  \caption{Impact of TATL factor on the NB cost $\left(C^{NB}\right)$ and
  different between investment cost of full $\mathcal{N}-1$ case and
  $(\mathcal{N}-1)'$ case with TATL
  $\left(C^{I(\mathcal{N}-1)}-C^{I(\mathcal{N}-1')}\right)$ in the sequential
  model in (a) 30\% (b) 90\% \coo\ emission reduction scenarios with
  $C_f=23$~\euro/kW/a.}
  \label{fig:tatl_seq}
\end{figure}

\begin{figure}[!t]
  \begin{subfigure}{0.5\linewidth}
    \centering
    \caption{$C_f=23$~\euro/kW/a}
    \includegraphics[width=\linewidth]{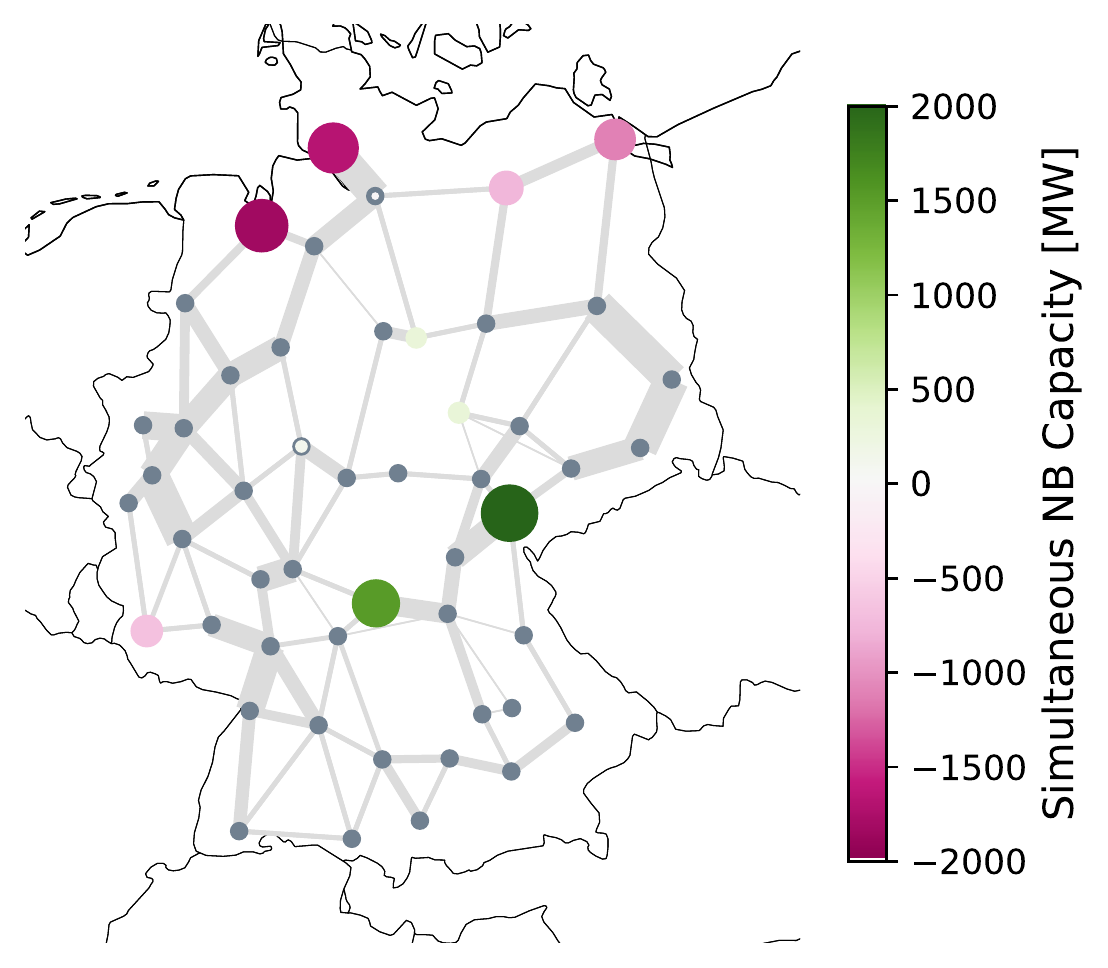}
  \end{subfigure}
  \begin{subfigure}{0.5\linewidth}
    \centering
    \caption{$C_f=0.010$~\euro/kW/a}
    \includegraphics[width=\linewidth]{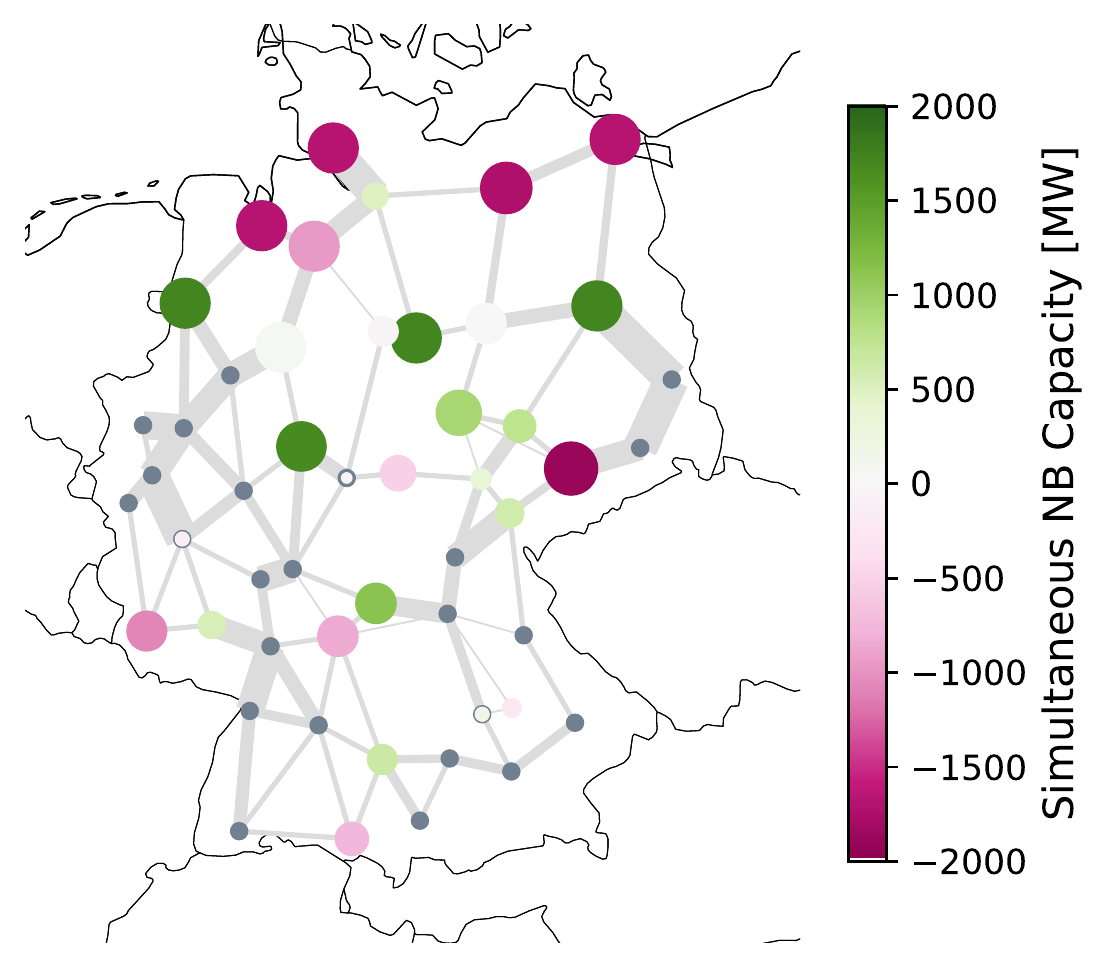}
  \end{subfigure}
  \caption{Impact of flexibility costs on allocated NB resources in the
  simultaneous model with $f^{tatl}=1.3$ in a 90\% \coo\ emission reduction
  scenario.}
  \label{fig:map_sim}
\end{figure}

In this section, the performance of the sequential model is assessed. In this
way, the impacts of different \coo\ emission reduction scenarios and the TATL
factor ($f^{tatl}$) on system costs and allocated flexibility are studied in the
power transmission system. Fig. \ref{fig:map_seq} illustrates the map of
allocated flexibility resources based on different \coo\ emission reduction
scenario in the German transmission network where the positive values represent
the discharge capacities $(P_{i}^{+})$ and negative values represent charge
capacities $(-P_{i}^{-})$ of the network boosters. At no node are both upward
$P_{i}^{+}$ and downward $P_{i}^{-}$ capacities installed. For a 90\% CO$_2$
reduction, most of the downward capacity is installed in the North of Germany,
to absorb power from the newly built wind farms, while the upward capacities are
built in the South to supply the power gap that could not be transported. This
reflects the dominant pattern of wind power flowing from North to South in the
German system. The total capacity of allocated flexibility resources in the 90\%
\coo\ emission reduction scenario is $15.8$~GW, which is $110\%$ higher than
total capacity of allocated flexibility resources in the 30\% \coo\ emission
reduction scenario with $f^{tatl}=1.3$, reflecting the higher wind share for the
tighter \coo\ target.

To evaluate this trend systematically, Fig. \ref{fig:carbon_seq}(a) shows the
sensitivity of varying the levels of \coo\ emission reduction on what is spent
on network booster infrastructure ($C^{NB}$) and the cost difference
($C^{I(\mathcal{N}-1)}-C^{I(\mathcal{N}-1')}$) between the generation
infrastructure plan of the full preventive case $\mathcal{N}-1$ (equivalent to
$f^{tatl}=1$) and the preventive case relaxed by the TATL $(\mathcal{N}-1)'$
with $f^{tatl}=1.3$ in the sequential model. As seen in Fig.
\ref{fig:carbon_seq}(a), $C^{NB}$ is less than the difference of
$C^{I(\mathcal{N}-1)}-C^{I(\mathcal{N}-1')}$ for \coo\ emission reduction
scenarios higher than $\approx 42\%$. In other words, investing in flexibility
resources to boost the network is beneficial compared to investing in
electricity generation even for moderate \coo\ emission reduction levels. Fig.
\ref{fig:carbon_seq}(b) shows the amount of energy produced from renewable
energy resources compared to fossil fuel generation.
%According to Fig. \ref{fig:carbon_seq}(b), the share of RES/fossil fuels is increased/decreased with increment of \coo\ emission reduction scenarios, and the amount of NB resources is increased slightly as well.

Additionally, the impact of varying the TATL factor on the NB cost and
difference between investment cost of full $\mathcal{N}-1$ and $\mathcal{N}-1'$
cases considering 30\% and 90\% \coo\ emission reduction scenarios in the
sequential model are illustrated in Figs. \ref{fig:tatl_seq} (a) and (b),
respectively. Raising the TATL factor increases both $C^{NB}$ and
$C^{I(\mathcal{N}-1)}-C^{I(\mathcal{N}-1')}$. The difference between
$C^{I(\mathcal{N}-1)}-C^{I(\mathcal{N}-1')}$ and $C^{NB}$ is increased with
rising TATL factors, until this difference stabilizes for TATL factors above
1.3. A similar result was also seen for the German grid in \citep
{kollenda2020}, where a TATL factor of 1.2 allows most of the benefits of an
infinite TATL to be obtained.

\subsection{Simultaneous model}

\begin{figure}[t]
  \centering
  \includegraphics[width=0.6\linewidth]{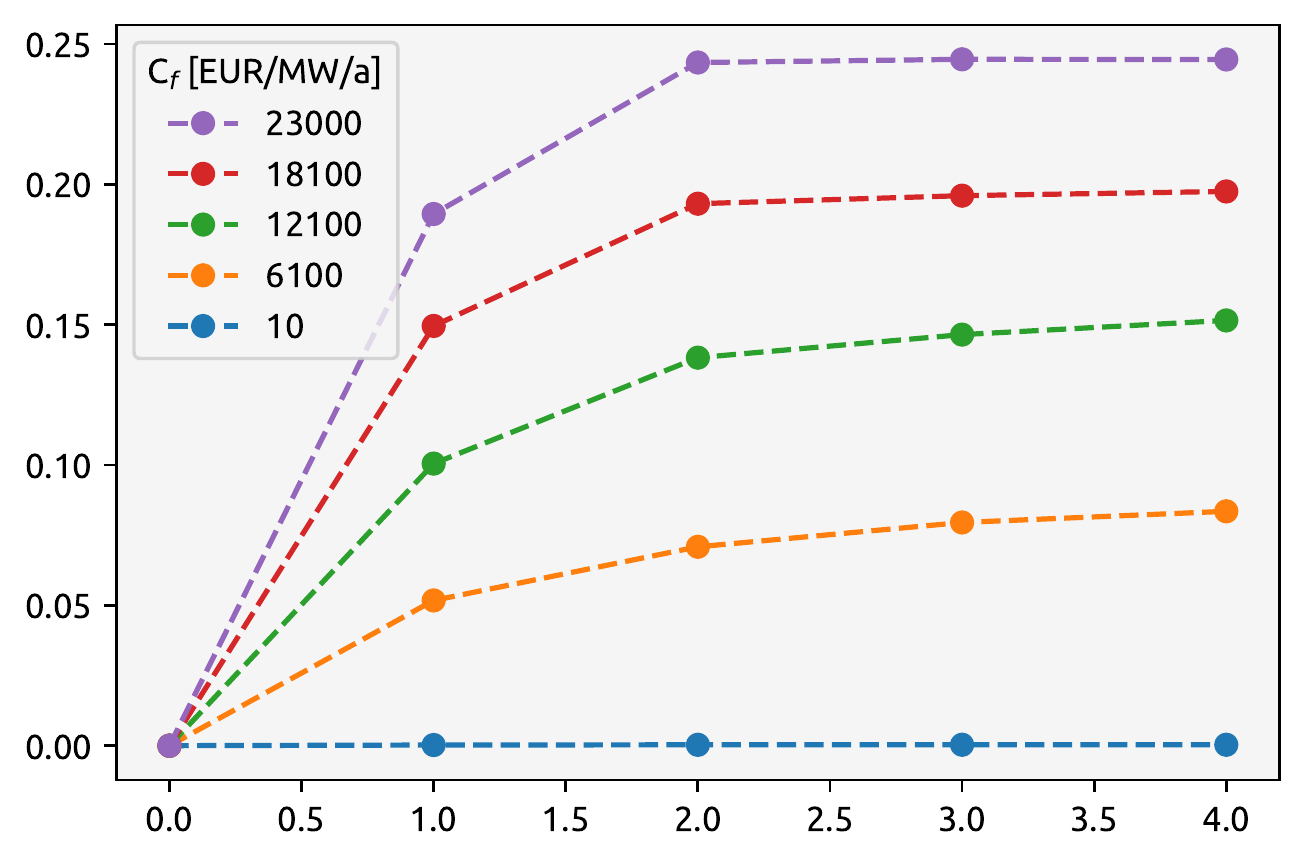}
  \caption{Impact of $f^{tatl}$ on the NB cost ($C^{NB}$) under various
  flexibility costs in the simultaneous model for a 90\% \coo\ emission
  reduction scenario.}
  \label{fig:tatl_sim}
\end{figure}

In this section, the performance of the proposed simultaneous model is
evaluated, while addressing sensitivities regarding \coo\ emission reduction
levels, flexibility cost ($C_f$), and the TATL factor. Fig. \ref{fig:map_sim}
displays a map of cost-optimally allocated NB infrastructure in a 90\% \coo\
emission reduction scenario for the highest and lowest considered flexibility
costs. With $C_f = 23$~\euro/kW/a in the 90\% \coo\ emission reduction scenario,
the total NB capacity drops from the 15.8~GW built in the sequential model to
10.5~GW for the simultaneous model, indicating that in the simultaneous case the
optimizer is choosing not to use corrective security for all line outages.
However, the figure shows that considerably more NB infrastructure is built if
flexibility costs are lower, with NB capacity rising from $10.5$~GW to $31$~GW.
As previously observed for the sequential model, power consuming network
boosters are predominantly found North of the power supplying network boosters
in both shown cases. Additionally, Fig. \ref{fig:tatl_sim} shows how the TATL
factor impacts the investment in network booster infrastructure for a 90\% \coo\
emission reduction scenario for various levels of flexibility costs. The figure
illustrates how increasing TATL factors make network boosters more attractive
across all flexibility cost levels. Analogous to the sequential model, the
investment in NB resources stabilises at TATL factors of 1.3 and beyond. The
total investment in network  booster infrastructure amounts to $0.24$ billion
Euros per year for the most conservative flexibility cost estimates.

\subsection{Cost benefits of corrective over preventive measures}

\begin{figure}[!t]
  \centering
  \includegraphics[width=0.6\linewidth]{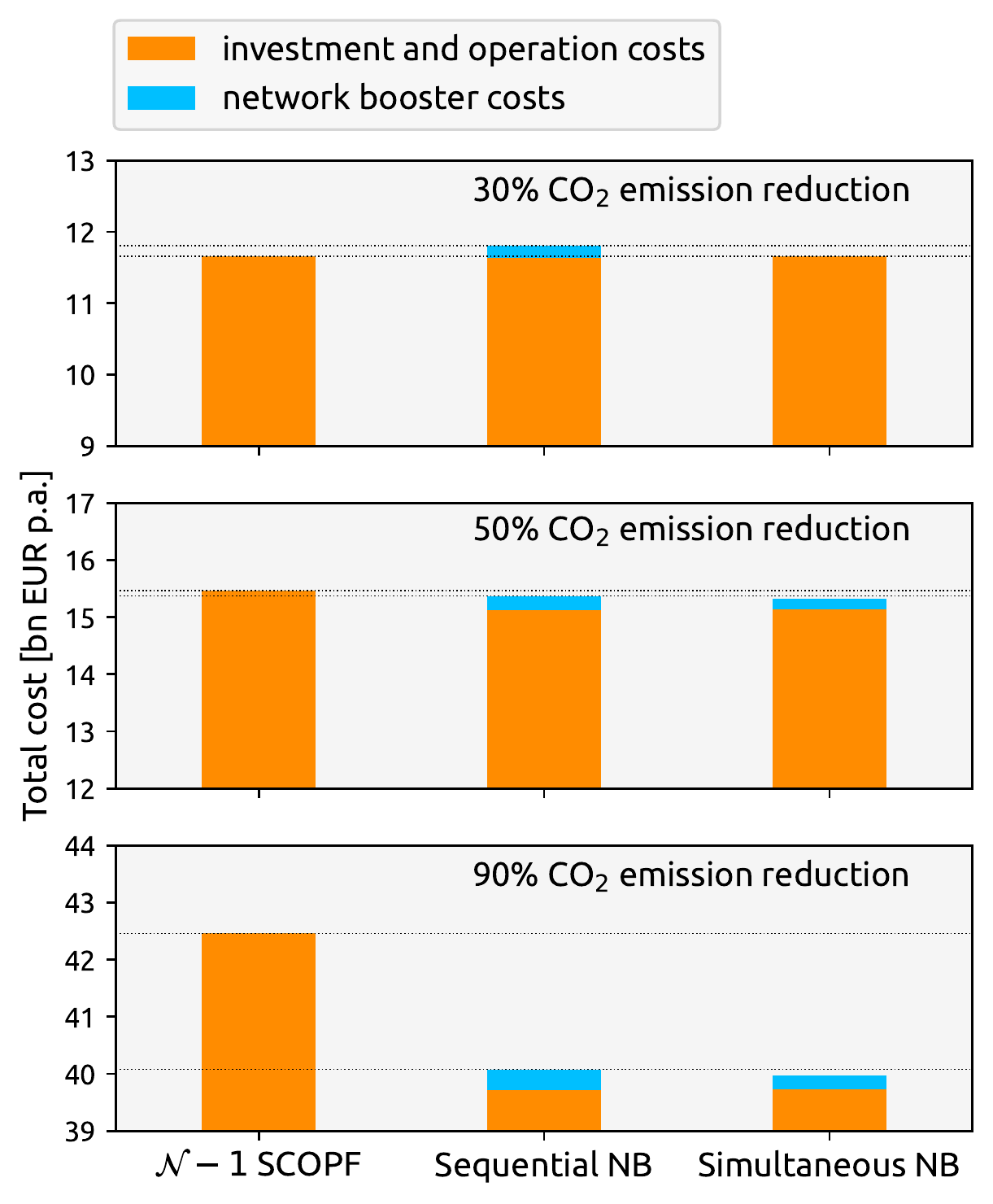}
  \caption{Comparison between total costs of the sequential and simultaneous
  strategies (with $f^{tatl}=1.3$ and $C^f=23$~\euro/kW/a) and SCOPF
  $\mathcal{N}-1$ for three different \coo\ emission reduction targets of 30\%
  (top), 50\% (center) and 90\% (bottom).}
  \label{fig:cost_seq_sim}
\end{figure}

In this section, the costs of the system based on sequential and simultaneous
models are compared to a system that is secured by standard preventive
$\mathcal{N}-1$ security constraints. Fig.~\ref{fig:cost_seq_sim} illustrates
the total system cost comprising electricity generation investment, operation,
and the NB costs for different \coo reduction targets.

The advantage of the simultaneous model is that it can choose for each line
failure a mixture of preventive and corrective measures. For a low \coo\ target
of 30\% there are low levels of wind and network loading, so the simultaneous
model does not build any network boosters and it is cost-optimal to apply a
preventative strategy everywhere equivalent to the standard  $\mathcal{N}-1$
case. For a target of 50\%, some network booster capacity appears in the
cost-optimal solution, but only 65\% of the total NB capacity of the sequential
model is built. Instead, a selective preventive-corrective strategy is applied
which sits between the standard $\mathcal{N}-1$ case and the full sequential NB
case. This results in costs which are below both extreme cases. Similarly, for a
90\% \coo\ reduction the simultaneous model chooses a mixed
preventative-corrective strategy with lower NB capacity than the sequential
model. However, the total system costs are more (or less) identical to the
sequential NB model which applies a corrective strategy for all line failures.
There is a substantial annual cost saving of $2.4$ billion euros from moving
from preventive to corrective security. There is an overall cost benefit of just
$0.25\%$ of the simultaneous model over the sequential model, which indicates
that there is only a small advantage to having a mixed preventive-corrective
approach compared to the purely corrective approach in the sequential model. In
addition, the sequential model has a clearer separation between generation,
planning and network security which may be advantageous from perspective of the
power network regulatory.

\section{Limitations and Future Work}\label{sec:limitations}

Our model contains several simplifications that are necessary to co-optimize
generation and network booster capacities within limited computational
resources. The bottleneck is the computer memory, since the simultaneous model
case study for Germany required 100~GB of RAM. Without these simplifications the
optimisation would not have been possible, but the simplifications may affect
the results. For example, the power flow has been linearized, but this neglects
reactive power flows and losses, which should be examined using the full
non-linear power flow equations. Unit commitment, discrete generation expansion
and intra-day uncertainty have been ignored. Similarly we have ignored
short-term dynamics that could result from the fast activation of the network
boosters, as well as the possible benefits of topology switching to deload the
network after a line failure. We have focused on the long-term benefits of the
network booster assuming  that the current single bidding zone for Germany with
subsequent nodal redispatch is replaced by a single market clearing process with
nodal power markets. There may be additional benefits for network boosters in
today's system where market dispatch is separated from congestion management. On
the other hand, planned network expansion may reduce the benefit from network
boosters. It is possible that network boosters can also be used for energy
arbitrage in the spot market or to provide reserves, adding possible revenue
streams and making the business case more attractive.

\section{Conclusion}\label{sec:conclusion}

In this paper, we have explored the long-term benefits of a corrective strategy
towards line failures compared to a traditional preventive security approach. We
have examined both a sequential model, where network boosters are assumed to
provide $\mathcal{N}-1$ security on all lines, and a simultaneous model which
allows mixed preventive-corrective strategies depending on local conditions.

According to our simulation results on a 50-bus German transmission network, we
found that using a corrective strategy frees up existing network capacities,
allowing cost-saving investments in high quality renewable sites that bring
total system cost savings of up to $2.4$ billion euro per year for a 90\% \coo\
reduction target. Even for lower targets, the selective use of network boosters
is beneficial. For a high share of renewable energy sources, a widespread use of
network boosters is optimal and there is only a small cost-benefit of the mixed
preventive-corrective strategy of the simultaneous model. Moreover, it is found
that once the temporarily admissible loading (TATL) rises 30\% beyond the
permanent rating, there is no additional cost saving, so that for lower TATL
most of the advantages of the network booster are still gained.

\section*{Acknowledgement}
This paper was conducted as part of the CoNDyNet2 project, which was supported
by the German Federal Ministry of Education and Research under grant number
03EK3055E.

\bibliographystyle{elsarticle-num.bst}
\bibliography{network-booster}

% \end{thebibliography}

% \begin{IEEEbiography}{Michael Shell} Biography text here. \end{IEEEbiography}

% % if you will not have a photo at all: \begin{IEEEbiographynophoto}{John Doe}
% Biography text here. \end{IEEEbiographynophoto}

% \begin{IEEEbiographynophoto}{Jane Doe} Biography text here.

\end{document}